\documentclass[12pt]{iopart}
\usepackage[latin9]{inputenc}
\usepackage{graphics}
\usepackage{graphicx}
\usepackage{amssymb}
\usepackage{epsfig}

\usepackage{xcolor} % Only for "\comment"

\usepackage{ucs}
\usepackage{bbm}

\newcommand{\meanO}[1]{\ensuremath{\langle #1 \rangle}}

\newcommand{\ket}[1]{|#1\rangle}
\newcommand{\ketbra}[1]{| #1\rangle \langle #1|}

\newcommand{\be}{\begin{equation}}
\newcommand{\ee}{\end{equation}}
\newcommand{\eea}{\end{eqnarray}}
\newcommand{\bea}{\begin{eqnarray}}

\newcommand{\va}[1]{\ensuremath{(\Delta#1)^2}}

\newcommand{\exs}[1]{\ensuremath{\langle{#1}\rangle}}

\newcommand{\kommentar}[1]{}
\newcommand{\trace}{{\rm Tr}}

\newcommand{\forget}[1]{}
\newcommand{\EQL}[1]{Equation~\eqref{#1}}
\newcommand{\EQLS}[1]{Equations~\eqref{#1}}

\newcommand{\SEC}[1]{section~\ref{#1}}
\newcommand{\FIG}[1]{figure~\ref{#1}}
\newcommand{\FIGL}[1]{Figure~\ref{#1}}
\newcommand{\REF}[1]{\cite{#1}}

\newcommand{\commentcolor}[1]{}

\newcommand{\binom}[2]{\bigg(\begin{array}{c} #1 \\
#2 \end{array}\bigg)}
\newcommand{\eqref}[1]{(\ref{#1})}
\newcommand{\EQ}[1]{(\ref{#1})}
\newcommand{\EQS}[1]{(\ref{#1})}

\newcommand{\tfrac}[2]{\frac{#1}{#2}}
\newcommand{\affiliation}[1]{\address{#1}}
\newcommand{\text}[1]{\mbox{#1}}

\newcommand{\komment}[1]{}

\begin{document}

\title{Detecting metrologically useful entanglement in the vicinity of Dicke states}

\author{Iagoba Apellaniz$^1$, Bernd L\"ucke$^2$, Jan Peise$^2$, Carsten Klempt$^2$, G\'eza T\'oth$^{1,3,4}$}
\affiliation{$^1$  Department of Theoretical Physics, University of the Basque Country
UPV/EHU, P.O. Box 644, E-48080 Bilbao, Spain}
%\email{iagoba.apellaniz@gmail.com}
\affiliation{$^2$  Institut f\"ur Quantenoptik, Leibniz Universit\"at Hannover, Welfengarten 1, 30167 Hannover, Germany}
\affiliation{$^3$ IKERBASQUE, Basque Foundation for Science, E-48013 Bilbao, Spain}
\affiliation{$^4$ Wigner Research Centre for Physics, Hungarian Academy of Sciences,
P.O. Box 49, H-1525 Budapest, Hungary}

%\author{Iagoba Apellaniz}
%\affiliation{Department of Theoretical Physics, University of the Basque Country
%UPV/EHU, P.O. Box 644, E-48080 Bilbao, Spain}
%\email{Iagoba.apellaniz@gmail.com}
%\author{Bernd L\"ucke}
%\affiliation{Institut f\"ur Theoretische Physik, Leibniz Universit\"at Hannover, Appelstra\ss e 2, D-30167 Hannover, Germany}
%\author{Jan Peise}
%\affiliation{Institut f\"ur Theoretische Physik, Leibniz Universit\"at Hannover, Appelstra\ss e 2, D-30167 Hannover, Germany}
%\author{Carsten Klempt}
%\affiliation{Institut f\"ur Theoretische Physik, Leibniz Universit\"at Hannover, Appelstra\ss e 2, D-30167 Hannover, Germany}
%\author{G\'eza T\'oth}
%\affiliation{IKERBASQUE, Basque Foundation for Science, E-48013 Bilbao, Spain}
%\affiliation{Wigner Research Centre for Physics, Hungarian Academy of Sciences,
%P.O. Box 49, H-1525 Budapest, Hungary}
%\email{toth@alumni.nd.edu}
%\homepage{http://www.gtoth.eu}

\eads{\mailto{iagoba.apellaniz@gmail.com}, \mailto{toth@alumni.nd.edu}}

%\pacs{03.67.Mn, 03.65.Ud, 42.50.St}


% 03.65.Ud Entanglement and quantum nonlocality
% 03.67.Mn Entanglement measures, witnesses, and other characterizations
% 42.50.St Nonclassical interferometry, subwavelength lithography
% 03.67.Bg Entanglement production and manipulation (for entanglement in Bose-Einstein condensates, see 03.75.Gg)
% 03.75.Gg	Entanglement and decoherence in Bose-Einstein condensates
% 03.75.Dg	Atom and neutron interferometry

%Our method determines the usefulness of the state for the protocol that estimates 
%the angle of rotation
%based on the measurement of the second moment of a total spin component, which is %especially suited for metrology with Dicke states.

\begin{abstract}
We present a method to verify the metrological usefulness of noisy Dicke states of a particle ensemble
with only a few collective measurements, without the need for a direct measurement of the sensitivity.
Our method determines the usefulness of the state for the usual protocol for estimating the angle of
rotation with Dicke states, which is based on the measurement of the second moment of a total spin
component. It can also be used to detect entangled states that are useful for quantum metrology. We
test our approach on recent experimental results.
\end{abstract}

\date{\today}

\maketitle

\section{Inroduction}

Quantum metrology is concerned with metrological tasks in which the quantumness
of the system plays an essential role 
\cite{Giovannetti19112004,doi:10.1142/S0219749909004839,2014arXiv1405.7703D,pezze2014quantum,Schaff2015Interferometry}. One of its key goals 
%of quantum metrology
is identifying  bounds for the highest precision achievable in parameter estimation tasks
in a quantum system, for example, by using the theory of quantum Fisher information \cite{helstrom1976quantum,holevo1982probabilistic,PhysRevLett.72.3439,braunstein1996generalized,petz2008quantum}. 
Recently, there has been a large effort connecting quantum metrology to quantum information science, in particular,
to the theory of quantum entanglement \cite{1751-8121-47-42-424006}.
It has turned out that, in linear interferometers, 
entanglement is needed to surpass the shot-noise limit corresponding to product states \cite{PhysRevLett.102.100401}.
It has been shown that fully entangled multipartite quantum states are needed to reach the maximal precision \cite{PhysRevA.85.022321,PhysRevA.85.022322,PhysRevA.82.012337}. In particular, the quantum Fisher information, a fundamental quantity in  metrology, can be used to detect multipartite entanglement.

After the theoretical findings mentioned above, it is crucial to know how large precision can be achieved in realistic, noisy systems  \cite{escher2011general,demkowicz2012elusive}. 
Thus, quantum metrology has been a driving force behind 
the numerous recent quantum optics experiments with cold gases and cold trapped ions,
which
were possible due to the rapid technological advancement in the field 
 \cite{leibfried2004toward,napolitano2011interaction,riedel2010atom,gross2010nonlinear}.
Quantum metrology played a central role even in the recent experiments  with the squeezed-light-enhanced gravitational wave detector GEO 600 \cite{PhysRevA.88.041802}.

There have been many experiments with fully polarized atomic ensembles 
in which the collective spin of the particles is rotated around an axis perpendicular to the mean spin (for instance by a homogeneous magnetic field) and the angle of the rotation is estimated based on collective measurements. 
It has also been verified experimentally that spin squeezing can result in a better precision compared to fully polarized product states (i.e., SU(2) coherent states) \cite{
% Review
0953-4075-45-10-103001,
% Cold atomic ensemble + light
% Spin Squeezed Atoms: A Macroscopic Entangled Ensemble Created by Light
PhysRevLett.83.1319,
% Spin Squeezing of Atomic Ensembles via Nuclear-Electronic Spin Entanglement
PhysRevLett.101.073601, 
% Quantum Noise Limited and Entanglement-Assisted Magnetometry
PhysRevLett.104.133601,
% Magnetic Sensitivity Beyond the Projection Noise Limit by Spin Squeezing
Sewell2012Magnetic,
% Mesoscopic atomic entanglement for precision measurements beyond the standard quantum limit
Appel2009Mesoscopic,
RevModPhys.82.1041,
% BEC
% Squeezing and entanglement in a Bose--Einstein condensate
esteve2008squeezing,
% Atom-chip-based generation of entanglement for quantum metrology
riedel2010atom,
% Quantum Metrology with a Scanning Probe Atom Interferometer
PhysRevLett.111.143001,
% Scalable Spin Squeezing for Quantum-Enhanced Magnetometry with Bose-Einstein Condensates
PhysRevLett.113.103004}
since spin-squeezed states are characterized by a reduced uncertainty in a direction orthogonal to the mean spin  
\cite{PhysRevA.47.5138,PhysRevA.50.67,PhysRevLett.86.4431,2011PhR...509...89M}.
A method has been presented for detecting metrologically useful entanglement for spin-squeezed states based on collective measurements \cite{PhysRevLett.102.100401}.

\begin{figure}
\begin{center}
\includegraphics[width=6.5cm]{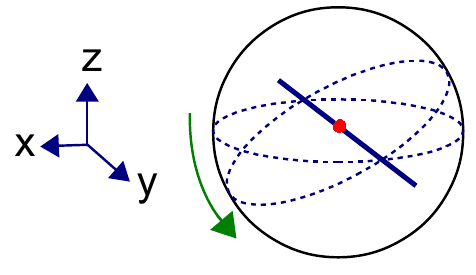} 
\caption{%(Color online) 
Metrology with symmetric Dicke states on the multiparticle Bloch sphere. In a  linear interferometer, the uncertainty ellipse of the state is rotated around the $y$ axis, while
the rotation angle is estimated by collective measurements.
}
\label{fig:metro}
\end{center}
\end{figure}

Besides almost fully polarized states, there are also unpolarized states considered for quantum metrology.
Prime examples of such states are Greenberger-Horne-Zelinger (GHZ) states \cite{ghz-argument}, which have already been realized experimentally many times 
\cite{ThreeQubitGHZ1,ThreeQubitGHZ,FourPhotonGHZ,lu2007experimental,gao2010experimental,leibfried2004toward,sackett2000experimental,PhysRevLett.106.130506}. Recently, new types of unpolarized states have been
 considered for metrology, such as singlet states \cite{PhysRevA.88.013626,2014arXiv1403.1964B} and symmetric 
Dicke states realized in cold gases and photons \cite{lucke2011twin,hamley2012spin,PhysRevLett.107.080504,PhysRevLett.103.020503}. 
 In the metrological schemes with Dicke states, the state is rotated around an axis in a linear interferometer and the rotation angle is estimated based on collective measurements (see \FIG{fig:metro}). 
 For this case, a criterion to detect the metrological usefulness of some of these states has been derived for symmetric systems \cite{1367-2630-16-10-103037}. There is another
criterion based on an improved Heisenberg relation to bound the quantum Fisher information close to Dicke
states \cite{frowis2014tighter}. However, these criteria show metrological usefulness allowing for arbitrary measurements using the
theory of the quantum Fisher information, while it might be interesting to show metrological usefulness for the
measurements carried out in a particular metrological scheme. For example, in a large ensemble, we can allow
for collective measurements only.

In this paper, we present a condition for metrological usefulness for the case when the second moment of a
spin component of the state is measured to obtain an estimate for the rotation angle. Our findings are expected
to simplify the experimental determination of metrological sensitivity since the proposed set of a few collective
measurements is much easier to carry out than determining the metrological sensitivity directly. Our method is
optimal in the sense that it gives the precision of the parameter estimation exactly, if certain operator expectation
values are provided. If not all relevant expectation values can be measured, it can still give useful bounds. We also
test our approach using data of a recent experiment realizing parameter estimation with a Dicke state \cite{PhysRevLett.112.155304}. Thus,
our paper is expected to be useful for similar experiments in the future. Since quantum states with a metrological
sensitivity larger than a certain bound are entangled \cite{PhysRevLett.102.100401}, our method can also be used to detect metrologically
useful entanglement in the vicinity of Dicke states. Note, however, that our work is not related to the criterion
presented in \REF{PhysRevLett.112.155304}, which detects multipartite entanglement in particle ensembles independently from
metrological applications.

Our paper is organized as follows. In \SEC{sec:basics}, we discuss the basics of quantum metrology. In \SEC{sec:Dicke}, we present our criterion.
 In \SEC{sec:Examples}, we compare our criterion to the sensitivity bound obtained from the quantum Fisher information.
 In \SEC{sec:Concrete}, we show how to apply our criterion to experimental results.
%In the appendix, we discuss some of the details of our calculations.

\section{Basics of quantum metrology}

\label{sec:basics}

In this section, we review the basics of quantum metrology. We discuss how the precision of the parameter estimation can be calculated, and how it can be bounded by the quantum Fisher information. We also discuss how the precision is linked to the entanglement of the quantum state.

One of the most fundamental tasks in quantum metrology is the estimation of
the small phase $\theta$ in the unitary dynamics 
\be\label{eq:rhotheta}
\varrho_\theta=e^{-iH\theta}\varrho e^{+iH\theta},
\ee
where $H$ is the Hamiltonian of the dynamics,
$\varrho$ is the initial state, and 
$\varrho_\theta$ is the final state after the evolution.
The parameter $\theta$ must be estimated based on measuring an observable $M$ on the final state. 

Next, we will discuss how to estimate the uncertainty of the parameter estimation. 
The variance of the estimated parameter  can be calculated by the error propagation formula as 
\be
(\Delta \theta)^{2}=\frac{(\Delta M)^2}{|\partial_{\theta} \langle{M}\rangle|^2}.\label{eq:errorprop}
\ee
One can interpret  \EQ{eq:errorprop} as follows. 
The larger the variance of $M,$ the worse the precision.
 On the other hand, the larger the derivative of the expectation value of $M,$ the better the precision. 

Some operators are better than others for 
parameter estimation with a certain quantum state. 
%One can even obtain the best precision achievable, allowing any operator $M$ %to be measured. 
The quantum Cram\'er-Rao inequality gives a lower bound on \EQ{eq:errorprop} 
that cannot be surpassed by any choice of $M.$
In this paper, we will use many times the  Cram\'er-Rao inequality
formulated with the reciprocal of $(\Delta \theta)^{2}$
as
\begin{equation} \label{eq:cr}
(\Delta \theta)^{-2}\le F_Q[\varrho,H],
\end{equation}
where 
%$(\Delta \theta)^{-2}$ is just the reciprocal of $(\Delta \theta)^{2}$
%and
$F_Q$ is the quantum Fisher information \cite{helstrom1976quantum,holevo1982probabilistic,PhysRevLett.72.3439,braunstein1996generalized}.
It has been shown that the bound  in \EQ{eq:cr} can be saturated by some measurement,
and there is even a formula to find the optimal observable \cite{PhysRevLett.72.3439}. Note that all these are valid in the limit
of infinite repetitions of the measurement, from
which the expectation values can be obtained exactly. The case of finite
number of measurements is more complicated \cite{2014arXiv1405.7703D,pezze2014quantum}.

In certain situations, it is better to use  \EQ{eq:errorprop} rather than  \EQ{eq:cr} for calculating the best precision achievable, since it gives the precision for a particular operator to be measured in an experimental setup.  This is reasonable since in a typical experiment, only a restricted set of operators can be measured. In this article, we will consider many-particle systems in which 
the particles cannot be addressed individually, and only collective quantities can be measured. In particular, in such a multiparticle system, we can measure the collective angular momentum operators
\be\label{eq:collective}
J_l=\sum_{n=1}^N j_l^{(n)},
\ee
for $l=x,y,z,$ and $j_l=\frac{1}{2}\sigma_l,$ where $\sigma_l$ are the Pauli spin matrices. Moreover,  $N$ is the number of pseudo-spin-$\frac{1}{2}$ particles.

Using collective angular momentum operators, it is even possible to connect the metrological precision to quantum entanglement \cite{PhysRevA.40.4277,RevModPhys.81.865}. Let us briefly review some notions of entanglement theory. Separable states are mixtures of multiparticle products states. 
If a state is not separable then it is called entangled.
 Entangled states can be used as a resource for several quantum information processing tasks \cite{RevModPhys.81.865}.
It has turned out that certain entangled states are also useful for quantum metrology.
In particular,  if a quantum state fullfils
\be \label{eq:FQN1}
F_Q[\varrho,J_l]>N,
\ee
then it is entangled \cite{PhysRevLett.102.100401}. As a consequence of  \EQ{eq:errorprop}, \EQ{eq:cr} and \EQ{eq:FQN1}, if 
\be\label{eq:FQN}
\frac{|\partial_{\theta} \langle{M}\rangle|^2}{(\Delta M)^2}>N
\ee
holds, then the system is also entangled.  Hence, entanglement is 
required for a large metrological precision.

Finally, it is even possible to find bounds for states with various forms of multipartite entanglement. Let us review very briefly the definitions needed to characterize multipartite entanglement. A pure state containing at most $k$-particle entanglement
is of the form
\be\label{eq:kpart}
\ket{\Psi_{k-{\rm particle\;ent.}}}=\ket{\psi_1}\otimes \ket{\psi_2} \otimes ... \otimes \ket{\psi_M},
\ee
where $\ket{\psi_k}$ are quantum states of at most $k$ qubits.  A mixed state containing at most $k$-particle entanglement is a mixture of pure states
of the form \eqref{eq:kpart}  \cite{PhysRevLett.86.4431,PhysRevLett.87.040401,1367-2630-7-1-229}. 
Recently, it has been shown that if  
\be\label{eq:FQmN}
F_Q[\varrho,J_l]>kN
\ee
holds for a quantum state, then it is at least $(k+1)$-particle entangled \cite{PhysRevA.85.022321,PhysRevA.85.022322}
\footnote[1]{Equation \eqref{eq:FQmN} is valid if $k$ is a divisor of $N.$
The general formula is somewhat more complicated than \EQ{eq:FQmN} \cite{PhysRevA.85.022321,PhysRevA.85.022322}.
Equation \eqref{eq:FQmN} can also be used if $k\ll N,$ since in this case
the difference between \eqref{eq:FQmN}  and the general formula is small. Note that $k\ll N$ is fulfilled in practice in many experiments. }. Another formulation is saying that the entanglement depth of the 
state is at least $(k+1)$ \cite{PhysRevLett.86.4431}.
Similarly to the previous paragraph, it follows that 
a quantum state is at least $(k+1)$-particle entangled if
\be\label{eq:FQmNF}
\frac{|\partial_{\theta} \langle{M}\rangle|^2}{(\Delta M)^2}>kN
\ee
holds.

Based on this section, one can see the advantages of using the quantity $(\Delta \theta)^{-2}$ rather than $(\Delta \theta)^{2}$ in our discussion.
It can be directly  compared to
the quantum Fisher information [see \EQ{eq:cr}].
Moreover, $(\Delta \theta)^{-2}/N$ directly leads to  a lower bound on the entanglement depth. 
%Hence, we will use 
%$(\Delta \theta)^{-2}$ throughout in this paper.
Note the relation of $(\Delta \theta)^{-2}$ to the precision: it is large for a high precision and small for a low precision.

\section{Metrology with Dicke states}
\label{sec:Dicke}

In this section, we will consider metrology with symmetric Dicke states \cite{PhysRev.93.99}. 
In particular, we will consider symmetric states that are the eigenstates of $J_z$ with a zero eigenvalue.
The metrological setup is the following. The Dicke state is rotated around the $y$ axis of the multiparticle Bloch sphere. Then, we estimate the rotation angle by collective measurements. Such an experiment has already been carried out in cold gases \cite{lucke2011twin}. 
It was found that for noisy states, the optimal angle for parameter estimation is not 
$\theta=0.$
Thus,  \EQ{eq:errorprop}  was recorded for many different values of $\theta.$
%they determined \EQ{eq:errorprop} for many $\theta$-values. 
%It was 
The phase estimation uncertainty was
then plotted as a function of the rotation angle $\theta,$
and the best precision could be identified.
In this section, we will show that the optimal angle can be obtained easily as a closed formula. 
We even find a closed formula for the maximal parameter estimation precision, as a function of a few expectation values. 
In this way, one can verify the metrological usefulness of the state without 
directly probing the phase estimation uncertainty for many phases.
%carrying out the metrology itself.

Next, let us define the Dicke states, and examine their metrological properties. 
A $N$-qubit symmetric Dicke state is given as
\begin{equation}\label{eq:DickeNm}
\ket{{\rm D}_N^{(m)}}=\binom{N}{m}^{-\frac{1}{2}}\sum_k \mathcal{P}_k (\ket{1}^{\otimes m}\otimes\ket{0}^{\otimes (N-m)}),
\end{equation}
where the summation is over all the different permutations of the product state having $m$ particles in the $\ket{1}$ state and $(N-m)$ particles in the $\ket{0}$ state.
One of such states is the $W$-state for which $m=1,$ which has been prepared with photons, ions, and neutral atoms \cite{PhysRevLett.92.077901,haffner2005scalable,Haas2014Entangled}. 

From the point of view of metrology, we are interested mostly in the symmetric Dicke state for even $N$ and $m=\frac{N}{2}.$ 
This state is known to be highly entangled \cite{Toth:07,1367-2630-11-8-083002} and allows for Heisenberg-limited interferometry \cite{PhysRevLett.71.1355}.
In the following, we will omit the superscript giving the number of  $\ket{1}$'s and use the notation
\begin{equation}\label{eq:Dicke}
\ket{{\rm D}_N}\equiv\ket{{\rm D}_N^{(\frac{N}{2})}}.
\end{equation}
Symmetric Dicke states of the type \eqref{eq:Dicke} have been created in photonic systems \cite{PhysRevLett.98.063604,PhysRevLett.103.020504,PhysRevLett.103.020503,PhysRevLett.107.080504,PhysRevLett.109.173604}, in cold gases 
\cite{lucke2011twin,hamley2012spin,PhysRevLett.112.155304} and recently in trapped cold ions \cite{Schindler2013Quantum}, 
and their metrological properties have also been verified experimentally \cite{lucke2011twin,PhysRevLett.107.080504}. 

How can we do metrology with a $\ket{{\rm D}_N}$ state, taking into account even the practical case of a nonideal Dicke state? 
We will consider a general initial state $\varrho,$ rather than the special case of a Dicke state.
We will study  a scheme in which the state is rotated around the $y$ axis, corresponding to a unitary evolution under the Hamiltonian 
\be\label{eq:HD}
H_{{\rm D}}=J_y.
\ee
Then, we measure $\meanO{J_z^2}$ to obtain an estimate for the angle of rotation.
The error propagation formula \EQ{eq:errorprop} gives us the variance of the parameter estimation as
\be
(\Delta \theta)^{2}=\frac{(\Delta J_z^2)^2}{|\partial_{\theta} 
\langle{J_z^2}\rangle|^2}.\label{eq:errorprop_Dicke}
\ee
Next, we calculate the quantites in \EQ{eq:errorprop_Dicke} one after the other.
For that, we need to use the dynamics of $J_z$ given in the Heisenberg picture as 
\be
J_z(\theta)=e^{iJ_y\theta}J_z e^{-iJ_y\theta}.\label{eq:dyn}
\ee
In the following, all operators evolve according to the Heisenberg picture
and  all expectation values are calculated for the initial state $\varrho.$

Before continuing our calculations, we need to make an important simplifying assumption. We will assume that for all  $\theta$
\bea
\langle{J_z^2(\theta)}\rangle&=&\langle{J_z^2(-\theta)}\rangle,\nonumber\\
\langle{J_z^4(\theta)}\rangle&=&\langle{J_z^4(-\theta)}\rangle
\label{eq:minustheta}
\eea
holds. 
\EQL{eq:minustheta} implies that the two expectation values must be even functions of $\theta,$
and that we can omit the terms
that are odd in $\theta.$ 
In \SEC{sec:Concrete}, we will see that unitary dynamics starting from the experimentally prepared state have the property \eqref{eq:minustheta}. 

Let us now continue computing the precision given by \EQ{eq:errorprop_Dicke}.
First, let us calculate the numerator of \EQ{eq:errorprop_Dicke}. Using \EQ{eq:dyn} to obtain the dynamics,
and with our simplifying assumptions  \eqref{eq:minustheta} we arrive at
\bea
\meanO{J_z^2 (\theta)} &=&
\meanO{J_z^2} \cos^2\theta + \meanO{J_x^2} \sin^2\theta, \nonumber\\
\meanO{J_z^4 (\theta)} &=& \meanO{J_z^4} \cos^4\theta + 
\meanO{J_x^4}\sin^4\theta \nonumber\\ &+& 
\bigg(\meanO{\{J_z,J_x\}^2}+\meanO{\{J_z^2,J_x^2\}}\bigg)  \label{eq:errorprop_den}
\cos^2\theta\sin^2\theta,
\eea
where $\{X,Y\}=XY+XY$ is the anticommutator of $X$ and $Y.$
Then, we calculate the denominator of
\EQ{eq:errorprop_Dicke}.
Using the dynamics of $\meanO{J_z^2 (\theta)}$ given in 
\EQ{eq:errorprop_den} and the assumption in \eqref{eq:minustheta}, we obtain the derivative as
\be
\partial_{\theta}\meanO{J_z^2}=2(\meanO{J_x^2}-\meanO{J_z^2})\cos\theta\sin\theta. \label{eq:errorprop_num}
\ee
The details of our calculations are given in appendix A.

Substituting \EQ{eq:errorprop_den} and \EQ{eq:errorprop_num} into the error propagation formula \EQ{eq:errorprop_Dicke}, after straightforward algebra, we arrive at
a simple expression for the parameter variance
\be
\fl (\Delta \theta)^{2}=\frac
{(\Delta J_x^2)^2 f(\theta)
+4\meanO{J_x^2}-3\meanO{J_y^2}-2\meanO{J_z^2}(1+\meanO{J_x^2})+
6\meanO{J_zJ_x^2J_z}}{4(\meanO{J_x^2}-\meanO{J_z^2})^2},\label{eq:mainformula2}
\ee
where
\be
f(\theta):=\left[
\tfrac{(\Delta J_z^2)^2}{(\Delta J_x^2)^2}
\tfrac{1}{\tan^2\theta}+\tan^2 \theta 
\right].\label{eq:mainformula2b}
\ee
For the details of the calculation, see appendix B.

Next, we determine the optimal angle $\theta$  that 
minimizes the parameter variance \eqref{eq:mainformula2}.
It is easy to see that the optimal angle has
to minimize also \EQ{eq:mainformula2b}.
The angle minimizing  \EQ{eq:mainformula2b} is given by
\be\label{eq:thetaopt}
\tan^2\theta_{\rm opt}=\sqrt{\tfrac{(\Delta J_z^2)^2}{(\Delta J_x^2)^2}}.
\ee
\EQL{eq:thetaopt} makes it possible to plan an experiment for the verification of the maximal accuracy such that we do not need to measure the sensitivity for a large range of $\theta$'s, but can 
target the parameter values close to the  optimal angle. 

Remarkably, we can even use \EQ{eq:thetaopt} to obtain an explicit formula from \EQ{eq:mainformula2} for the minimal parameter variance achievable by the setup as
\be
\fl (\Delta \theta)^{2}_{\rm opt}=\frac
{2\sqrt{(\Delta J_z^2)^2(\Delta J_x^2)^2}
+4\meanO{J_x^2}-3\meanO{J_y^2}
-2\meanO{J_z^2}(1+\meanO{J_x^2})
+6\meanO{J_zJ_x^2J_z}}{4(\meanO{J_x^2}-\meanO{J_z^2})^2}.
\label{eq:mainformula}
\ee
For the evaluation of \EQ{eq:mainformula}, we do not need to make a direct measurement of the sensitivity for some range of $\theta$ in the  vicinity of $\theta_{\rm opt}.$ 
We need 
 to measure only  the expectation values $\exs{J_x^2},$ $\exs{J_y^2},$  $\exs{J_z^2},$ $\exs{J_x^4},$ $\exs{J_z^4},$ and 
$\exs{J_zJ_x^2J_z}$ of the initial state (i.e., at $\theta=0$),
which could make the  experiments much easier.
 Later, we will discuss how to avoid measuring $\exs{J_zJ_x^2J_z},$ and even avoiding
measuring the fourth order moments.

Finally, let us demonstrate the correctness of our formula \eqref{eq:mainformula} for the pure Dicke state $\ket{{\rm D}_N}.$ 
For this purpose, we will summarize the expectation values of the relevant moments of some collective observables for the state.
Our Dicke state is 
 an eigenstate of $J_z$ with an eigenvalue zero. Hence, it   immediately follows that
 \begin{equation}
\exs{J_z^2}=0, \;\;\; \;\;\; \;\;\; \;\;\; \exs{J_z^4}=0,\;\;\; \;\;\; \;\;\;  \;\;\;\exs{J_zJ_x^2J_z}=0 .\label{eq:paramvalues1}
\end{equation}
Moreover we know that for every quantum state 
\be\label{eq:paramvalues2}
\exs{J_x^2}+\exs{J_y^2}+\exs{J_z^2}\le \frac{N(N+2)}{4}
\ee
holds, while symmetric quantum states, such as the Dicke state $\ket{{\rm D}_N},$ saturate the inequality.
Based on \EQ{eq:paramvalues1} and \EQ{eq:paramvalues2},
and knowing that the rotational symmetry around the $z$ axis implies $\exs{J_x^2}=\exs{J_y^2},$ we arrive at 
\be\label{eq:paramvalues3}
\exs{J_x^2}=\exs{J_y^2}=\frac{N(N+2)}{8} .%\approx\frac{1}{2}\left(\frac{N}{2}\right)^2.
\ee
Somewhat technical, but straighforward algebra leads to 
\be\label{eq:paramvalues3b}
\exs{J_x^4}=\exs{J_y^4}=\frac{(N+2)}{8}\left(\frac{3N(N+2)}{16}-\frac{1}{2}\right),
%\approx\frac{3}{2}\left(\frac{N}{2}\right)^4.
\ee
which will be useful later in the article.
\EQLS{eq:paramvalues1} and \eqref{eq:paramvalues3} are sufficient to evaluate \EQ{eq:mainformula},
and we obtain
\bea
(\Delta \theta)^{2}_{\rm opt}&=&\frac{2}{N(N+2)},
\eea
which reproduces the  value 
given by the quantum Fisher information \cite{lucke2011twin}.
Hence, for this case the Cram\'er-Rao bound  \eqref{eq:cr} is saturated, which 
also means that $J_z^2$ is the optimal operator
to measure for the ideal Dicke state. In addition,  \EQ{eq:thetaopt} yields that the optimal angle
for the ideal Dicke state \eqref{eq:Dicke} is $\theta_{\rm opt}=0.$

\section{Testing our bound on concrete examples}
\label{sec:Examples}

In this section, we compare our formula  \eqref{eq:mainformula}  for $\va{\theta}_{\rm opt}$ with the bound obtained from the quantum Fisher information.
We find that the formula gives a good lower bound on the quantum Fisher information
using the inequality $(\Delta \theta)^{-2}_{\rm opt}\le F_Q[\varrho,J_y].$
It has been mentioned in the introduction that our formula yields the best precision assuming that $\exs{J_z^2}$ is measured 
after the linear interferometer. If a different operator is measured, then the precision can even be higher. The quantum Fisher information gives us a bound on the precision if any measurement is allowed. However, note that in the latter case the optimal measurement might turn out to be impractical.

\begin{figure}
\begin{center}
\includegraphics[width=6.65cm]{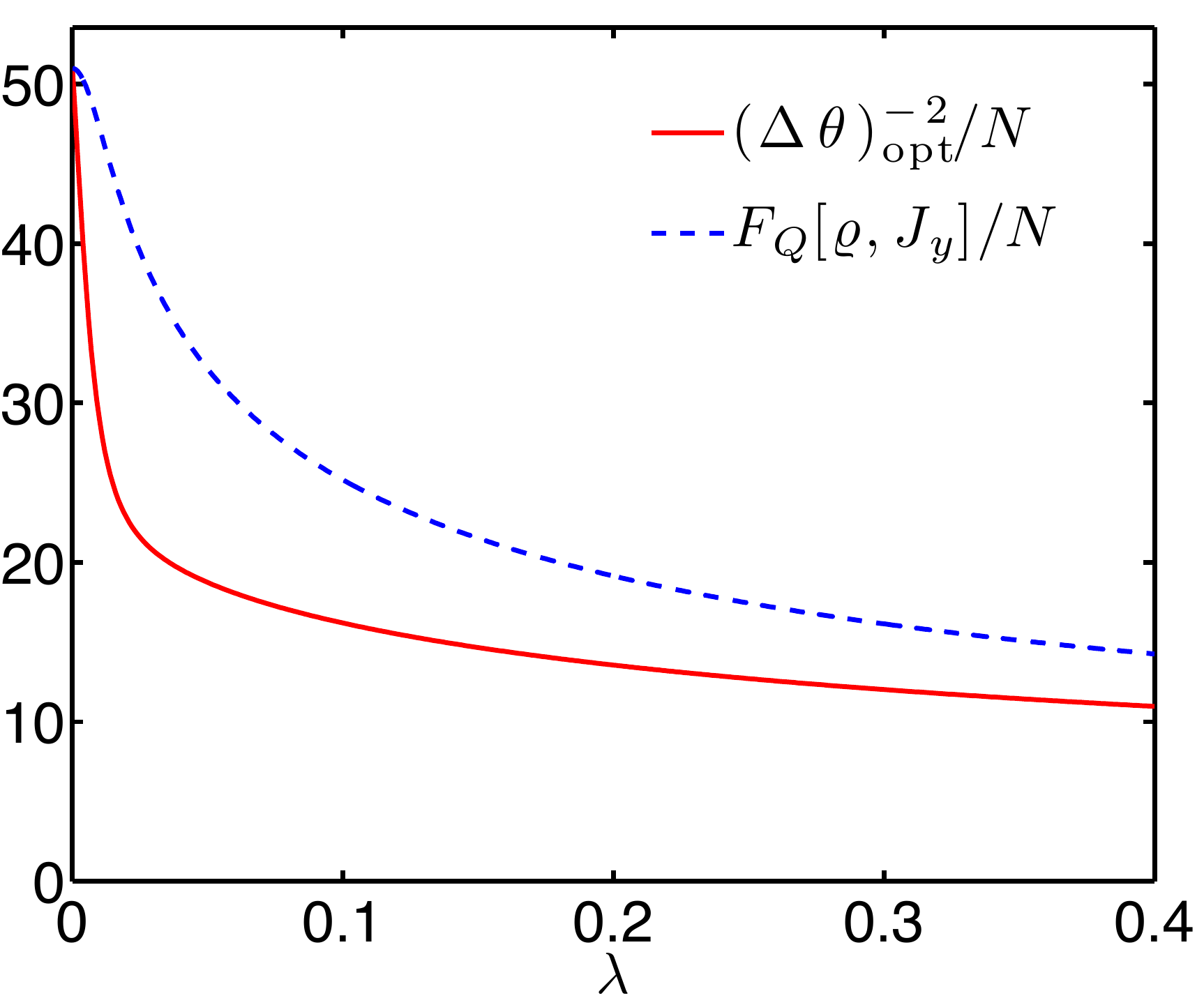} \includegraphics[width=6.5cm]{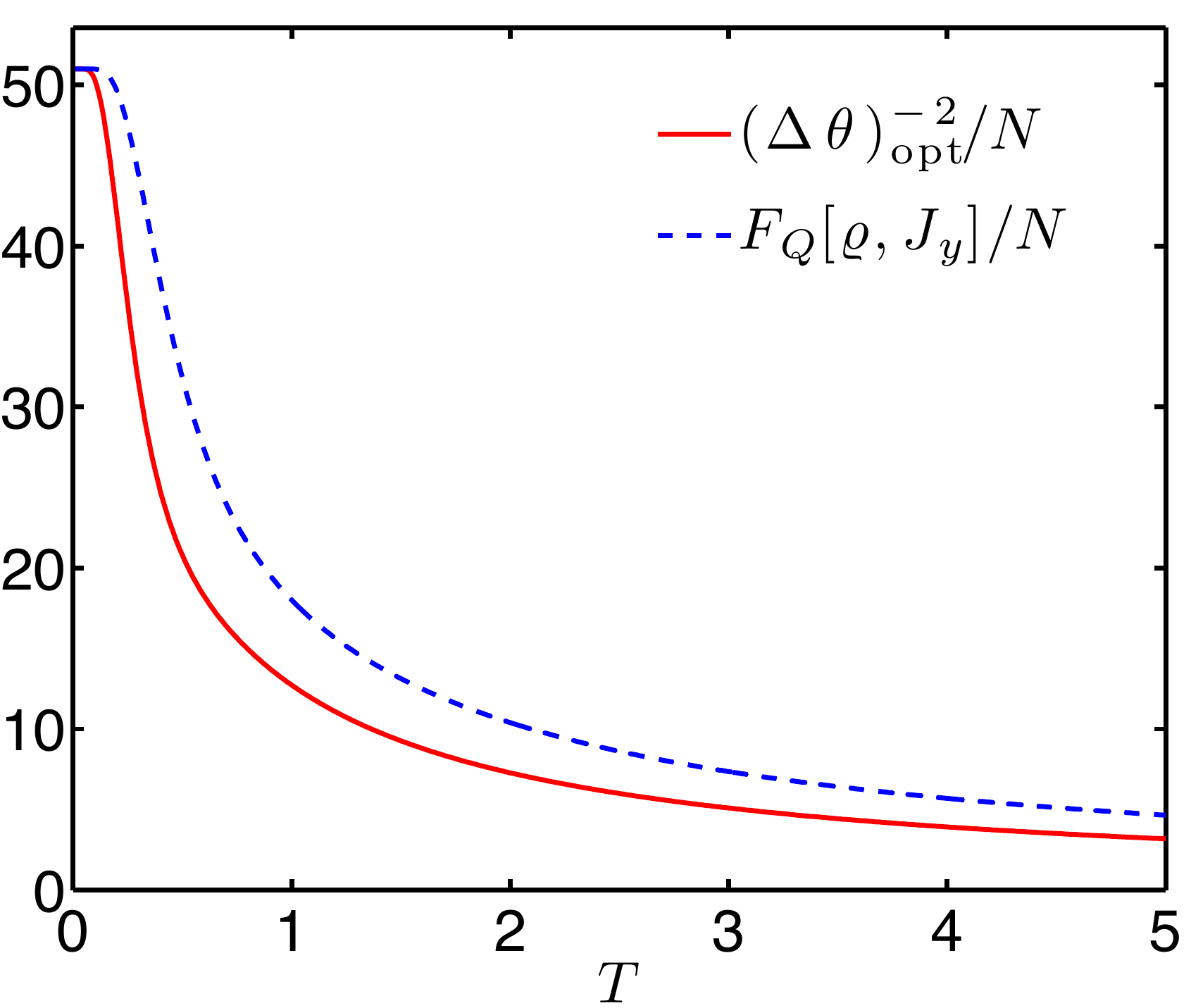} 

\hskip0.3cm (a) \hskip6.1cm (b)

\caption{%(Color online) 
(a) The reciprocal of the parameter variance $(\Delta \theta)^{2}_{\rm opt}$ given in \eqref{eq:mainformula} and the quantum Fisher information for the ground states of the spin-squeezing Hamiltonian \eqref{eq:Hsq} as a function of $\lambda$ for $N=100$ particles.
(b) The same quantities are shown for the thermal states  \eqref{eq:HT} as a function of $T.$
}
\label{fig:example1}
\end{center}
\end{figure}

Let us consider first the example of pure spin-squeezed states obtained as a ground state of the 
spin squeezing Hamiltonian
\be\label{eq:Hsq}
H_{\rm sq}(\lambda)=J_z^2-\lambda J_x,
\ee
where $\lambda$ is a real parameter. For  $\lambda>0,$ the ground state is unique, and it is in the symmetric 
subspace. Hence, we can use the SU(2) generators instead of the collective operators $J_l$ defined in \EQ{eq:collective} \cite{PhysRevLett.86.4431}.
We will get the same result, however, we can model larger systems this way.
For $\lambda\rightarrow \infty,$ the ground state is the fully polarized state in the $x$-direction.
For $\lambda\rightarrow +0,$ it is the Dicke state \EQ{eq:Dicke}. For intermediate $\lambda$ values, the ground state is a state which is polarized in the $x$-direction and spin squeezed in the $z$-direction.
We will now find the best precision that can be achieved with this state if we consider 
estimating $\theta$ in the unitary dynamics
\be\label{eq:unidyny}
\varrho_\theta=e^{-iJ_y\theta} \varrho_0 e^{+iJ_y\theta}.
\ee
 \FIGL{fig:example1}(a) compares the sensitivity we obtained with the optimum defined by the quantum Fisher information.
Our bound is close to the optimum 
when the state is well polarized. It also coincides with the bound in the $\lambda\rightarrow0$ limit, when the ground state is close to a Dicke state.

Our next example is a noisy Dicke state of the form
\be\label{eq:HT}
\varrho_{\rm th}(T)\propto \sum_{m=0}^{N} e^{-\frac{(m-N/2)^2}{T}}\ketbra{{\rm D}_N^{(m)}},
\ee 
where $N$ is even and $\ket{{\rm D}_N^{(m)}}$ is defined in \EQ{eq:DickeNm}.
From \EQ{eq:HT}, we obtain the Dicke state  $\ket{{\rm D}_N^{(N/2)}}$ for $T=0.$ For $T>0,$ other symmetric Dicke states in the vicinity of the state  
$\ket{{\rm D}_N^{(N/2)}}$ are also populated.
The distribution of Dicke states is Gaussian and \EQ{eq:HT} can be interpreted as a thermal state.
We consider again estimating the parameter $\theta$ in the dynamics \eqref{eq:unidyny}.
The results can be seen in  \FIG{fig:example1}(b). Again, our bound is quite close to ultimate bound 
defined by the  quantum Fisher information.

Next, we verify that the dynamics fulfill the condition \eqref{eq:minustheta} for both cases considered in this section. In this way we demonstrate that it was justified to use the formula \eqref{eq:mainformula} to obtain the precision. Simple algebra shows that 
if the states considered in our examples are used for metrology as initial states
then 
\be\label{eq:Jz_plus_minus_theta}
\trace(e^{-iJ_y\phi}\varrho e^{+iJ_y\phi}J_z^m)=\trace(e^{+iJ_y\phi}\varrho e^{-iJ_y\phi}J_z^m)
\ee
holds for $m=2,4,$ from which \EQ{eq:minustheta} follows.

Finally, note that in  \FIG{fig:example1}(a) and \FIG{fig:example1}(b)
a metrologically useful $(k+1)-$particle entanglement is detected based on \EQ{eq:FQmNF} if the quantum Fisher information divided by $N$ is larger than an integer  $k$ \footnote[2]{This is true if $k$ is a divisor of $N,$ or $k\ll N$ \cite{PhysRevA.85.022321,PhysRevA.85.022322}.}. Based on \EQ{eq:FQmNF}, a similar statement holds  for 
$(\Delta \theta)^{-2}_{\rm opt}/N,$ which detects entanglement that is useful for the metrological procedure with a $\exs{J_z^2}$ measurement.

\section{Applications of the method to experimental data}
\label{sec:Concrete}

In this section, we discuss how to apply the formula \eqref{eq:mainformula} in the cold gas experiment described in \REF{PhysRevLett.112.155304}.
In the experiment, it is possible to measure the operator $J_z,$ which is defined as a population difference as
\be
J_z=\frac{1}{2}(N_{+1/2}-N_{-1/2}),
\ee
where $N_{+1/2}$ and $N_{-1/2}$ are the number of particles in the spin-states $j_z=+1/2$ and $j_z=-1/2,$ respectively.
Hence, in principle the expectation values of all moments of $J_z$ can be obtained.
In practice, it is possible to measure the lower order moments like $\exs{J_z^2}$ and $\exs{J_z^4},$
while higher order moments necessitate an increasing number of repetitions of the experiment to get sufficient statistics.

The angular momentum components $J_x$ and $J_y$ are measured by rotating the total spin using a $\frac{\pi}{2}$ microwave coupling pulse before the 
population difference measurement. Whether $J_x$ or $J_y$ is measured depends on the relation between the microwave phase and the phase of the initial Bose-Einstein condensate. The condensate phase represents the only possible phase reference in analogy to the local oscillator in optics. Intrinsically, it has no relation to the microwave phase, such that we homogeneously average over all possible phase relations in our measurements. 
From another point of view, one can also say that the fluctuation of the phase results in a random rotation of the spin around the $z$ axis. Hence, we measure 
\be 
J_\alpha=\sin(\alpha)J_x+\cos(\alpha)J_y,
\ee
where $\alpha$ is a random phase, and we need to consider an averaging 
over $\alpha.$ 
Effectively, the density matrix of the state is
\be\label{eq:int}
\varrho=\frac{1}{2\pi} \int e^{-iJ_z\phi}\varrho_0 e^{+iJ_z\phi} {\rm d}\phi,
\ee
where $\varrho_0$ is what we would obtain if we had  access to the phase reference 
\footnote[3]{States of the form \eqref{eq:int} can be written
as incoherent mixtures of states with a definite $J_z,$ i.e.,
$\varrho=\sum_{l=-N/2}^{N/2} p_l \varrho_l,$ 
where for the subensembles
$\exs{J_z}_{\varrho_l}=l$ and $\va{J_z}_{\varrho_l}=0$ hold, while
for the probabilities
$p_l\ge 0$ and $\sum_l p_l=1$.}.
For a state of the form \eqref{eq:int}, the equality \eqref{eq:Jz_plus_minus_theta}
holds for $m=2,4,$ which can be seen directly by substituting \EQ{eq:int} into \EQ{eq:Jz_plus_minus_theta}.
Hence, the unitary dynamics will fulfill the simplifying assumption \eqref{eq:minustheta}.
Note that integration over the rotation angle in \EQ{eq:int} does not create quantum entanglement.
If the state $\varrho$ is entangled, $\varrho_0$ must also be entangled. 

Next, we will simplify the bound for the precision
of the parameter estimation  \eqref{eq:mainformula}, based on the
consequences of our state having the form \eqref{eq:int}.
Since $\varrho$ is invariant under rotations around the $z$  axis, we have
\be\label{eq:Jxym}
\exs{J_\alpha^m}=\exs{J_x^m}=\exs{J_y^m},
\ee
for all $m.$
Hence, the expectation values $\exs{J_x^m}$ and $\exs{J_y^m}$ can be obtained from measurements of 
$\exs{J_\alpha^m}.$ 
Moreover, there is a single remaining term in \EQ{eq:mainformula}, the expectation value $\meanO{J_z J_x^2 J_z},$  which is difficult to measure directly in an experiment. It can be bounded as
\bea
\meanO{J_z J_x^2 J_z}&=&\frac{\meanO{J_z (J_x^2+J_y^2) J_z}}{2}
=\frac{\meanO{J_z (J_x^2+J_y^2+J_z^2) J_z} - \meanO{J_z^4}}{2}
\nonumber\\&\le& \tfrac{N(N+2)}{8}\meanO{J_z^2}- \tfrac{1}{2}\meanO{J_z^4}=:\mathcal{Z},\label{eq:JzJx2Jzbound}
\eea
where the last inequality is due to \EQ{eq:paramvalues2}, which is saturated for symmetric states.
Thus, for symmetric states  the formula \eqref{eq:JzJx2Jzbound}
is not only an upper bound, it is exact.
Using \EQ{eq:Jxym} and \EQ{eq:JzJx2Jzbound}, we can simplify
\eqref{eq:mainformula} as
\be
(\Delta \theta)^{2}_{\rm opt}\le\frac
{2\sqrt{(\Delta J_z^2)^2(\Delta J_x^2)^2}
+\meanO{J^2_x}
-2\meanO{J_z^2}(1+\meanO{J^2_x})
+6\mathcal{Z}}{4(\meanO{J^2_x}-\meanO{J_z^2})^2},\label{eq:estimatedeltaphi}
\ee
where $\mathcal{Z}$ is defined in  \EQ{eq:JzJx2Jzbound}.

Next, we will substitute the experimentally measured values to \EQ{eq:estimatedeltaphi}. The measured data from \REF{PhysRevLett.107.080504} for $N=7900$ yields 
\begin{eqnarray}
\exs{J_z^2}=112\pm31, \;\;\; \;\;\;\;\;\; \;\;\; \;\;\; \;\;\; \;\;\; \;\;\;\;  \exs{J_z^4}=40\times10^3 \pm22\times10^3,\nonumber\\
\exs{J_x^2}=6\times10^6\pm 0.6 \times 10^6, \;\;\; \;\;\;\;\; \;\exs{J_x^4}= 6.2\times10^{13}\pm 0.8\times10^{13}.\label{eq:paramvalues}
\end{eqnarray}
Hence, we obtain for the precision
\be
\label{eq:result}
\frac{(\Delta \theta)^{-2}_{\rm opt}}{N} \ge 3.7 \pm 1.5.
\ee
In \EQS{eq:paramvalues} and \eqref{eq:result}, the statistical uncertainties have been obtained through boot strapping.
Note that a direct 
substitution of the mean values into \EQ{eq:estimatedeltaphi} would yield a gain of $3.3.$
Based on \EQ{eq:FQN}, \EQ{eq:result} proves the presence of metrologically useful entanglement \cite{PhysRevLett.102.100401}.
Based on \EQ{eq:FQmN}, it even indicates that the quantum state had metrologically useful $4$-particle entanglement.
Within one standard deviation, it demonstrates  $3$-particle entanglement.

\begin{figure}
\begin{center}
\includegraphics[width=8cm]{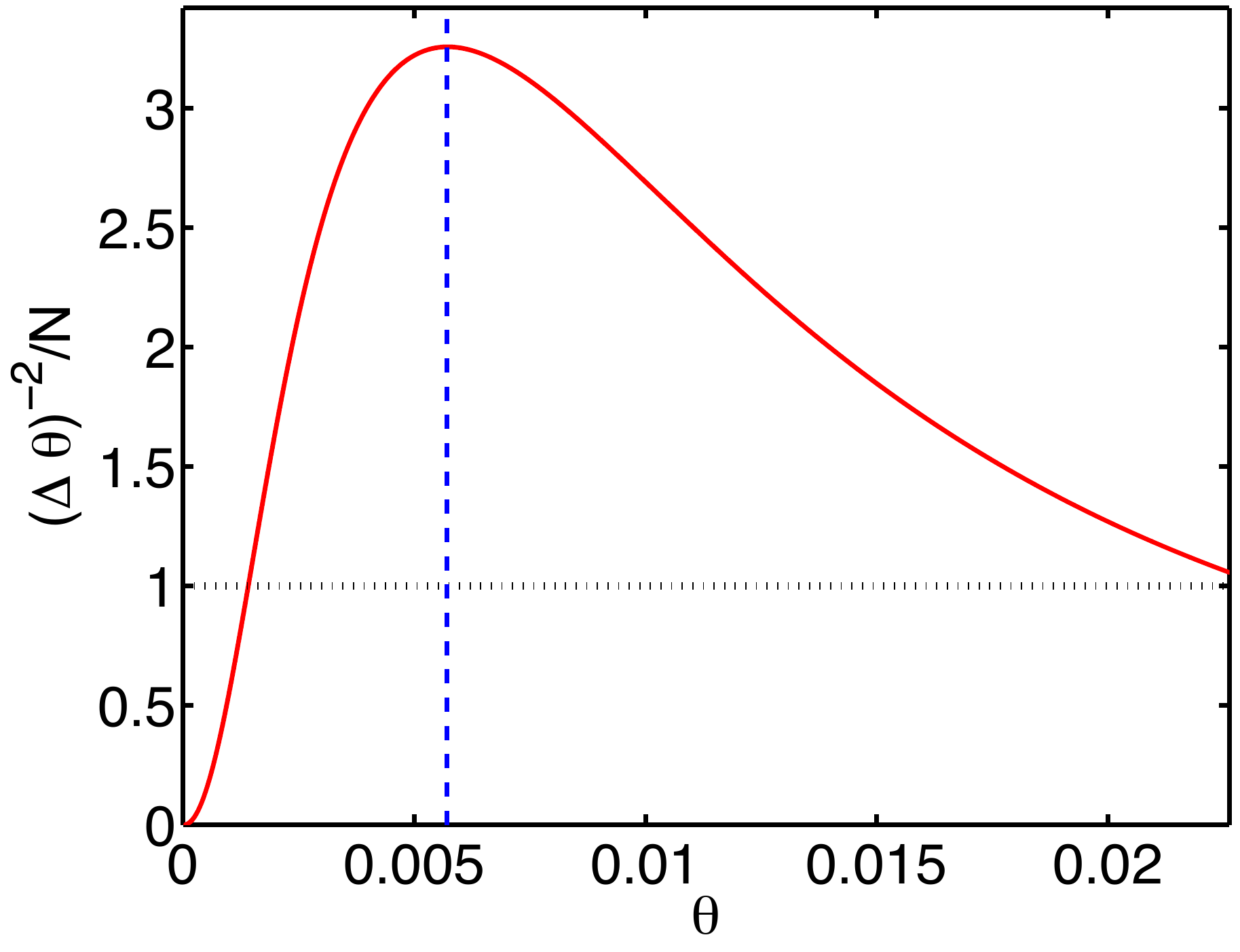} 
\caption{%(Color online) 
(solid) $(\Delta \theta)^{-2}/N$ as a function of the parameter $\theta$ given by \EQ{eq:mainformula2}, for parameter values given in \EQ{eq:paramvalues}. (dashed) The maximum is taken at $\theta=0.0057,$ as calculated based on \EQ{eq:thetaopt}. (dotted) $(\Delta \theta)^{-2}/N$ corresponding to the shot-noise limit.
If  the curve is above this line, then the quantum state shows entanglement based on \EQ{eq:FQN}. }
\label{fig:curve}
\end{center}
\end{figure}

In \FIG{fig:curve}, we plot the precision as a function of the rotation angle using the expectation values   
\eqref{eq:paramvalues} obtained experimentally. Since we cannot obtain the expectation value $\meanO{J_z J_x^2 J_z}$  experimentally,
we approximate it with the right-hand side of 
\EQ{eq:JzJx2Jzbound}, i.e., we plot the right-hand side of \EQ{eq:estimatedeltaphi}.
 With that, we overestimate $(\Delta \theta)^{2},$
or equivalently we underestimate $(\Delta \theta)^{-2}.$

Thus, we could detect metrological usefulness by measuring the second and fourth moments of the collective angular momentum components.
 For future applications of our scheme, it is important to further  reduce the number of quantities 
 we need to measure for our method.
In practice, one can easily avoid the need for determining $\meanO{J_x^4}$.
Note that the distribution of values obtained from measuring $J_x$ is strongly non-Gaussian. 
The values $\pm N/2$ appear most frequently,
and the value $0$ appear least frequently \cite{lucke2011twin}.
One can bound the fourth moment of $J_x$ as follows
\be\label{eq:approx}
\meanO{J_x^4}\leq\frac{N^2}{4}\meanO{J_x^2}.
\ee
\EQL{eq:approx} is based on the fact that for two commuting positive-semidefinite observables, $A$ and $B,$ we have
\be\label{eq:approx_matrix}
\meanO{AB}\leq\lambda_{\rm max}(A)\meanO{B},
\ee
where $\lambda_{\rm max}(A)$ is the largest eigenvalue of $A.$
Since even for a noisy Dicke state $\meanO{J_x^2}$ is very large,  \EQ{eq:approx} is
a very good upper bound. Substituting the right-hand side of  \EQ{eq:approx} in the place of $\meanO{J_x^4}$ into 
\EQ{eq:estimatedeltaphi}, we will underestimate $(\Delta \theta)^{-2}.$

It is also possible to approximate $\meanO{J_z^4}$ with $\meanO{J_z^2}.$
This will not lead to a strict bound on the precision as the one for $\meanO{J_x^2},$
but still can help us to access the metrological usefulness based on second moments only.
One can use the formula
\be\label{eq:approxz}
\meanO{J_z^4}=\beta\meanO{J_z^2}^2,
\ee
where $\beta$ is a constant. In principle, $\beta$ can be obtained based on
some knowledge of the distribution of the measured values.
In practice, the distribution is typically dominated by a Gaussian technical noise.
 For a Gaussian distribution and for large $N,$ we have $\beta=3.$ 
Note that the distribution is expected to be centered around zero, since the method used to 
create a Dicke state makes sure that $\exs{J_z}=0$ \cite{lucke2011twin,PhysRevLett.107.080504}.
Thus, \EQ{eq:approxz} can give an estimate on the fourth moment, even
if only the second moments are measured, 
under the assumption of a Gaussian probability density.
%and we assumed a Gaussian
%probability density of the measured values.

\begin{figure}
\begin{center}
\includegraphics[width=6.3cm]{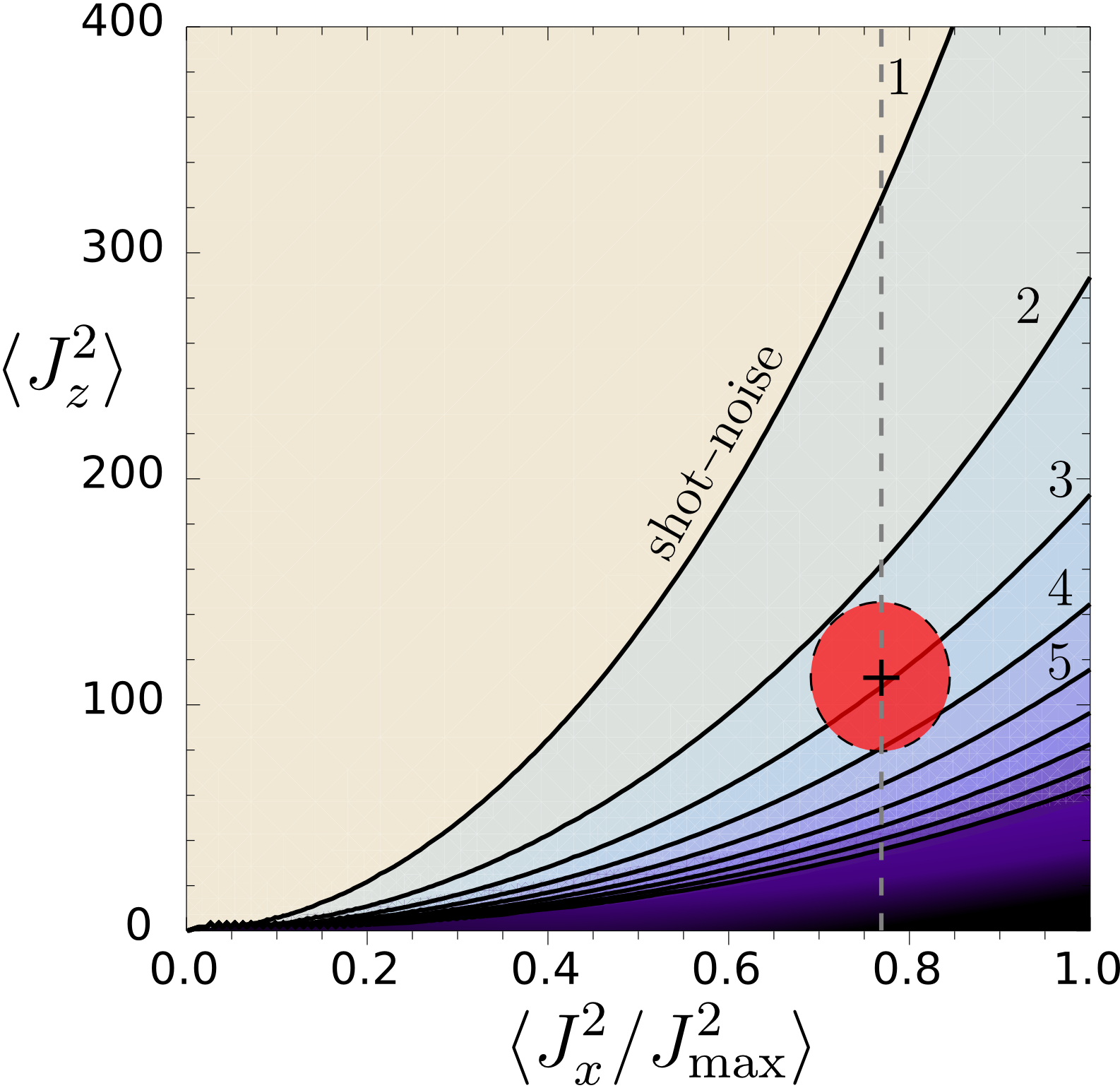} \hskip0.5cm 
\includegraphics[width=6.7cm]{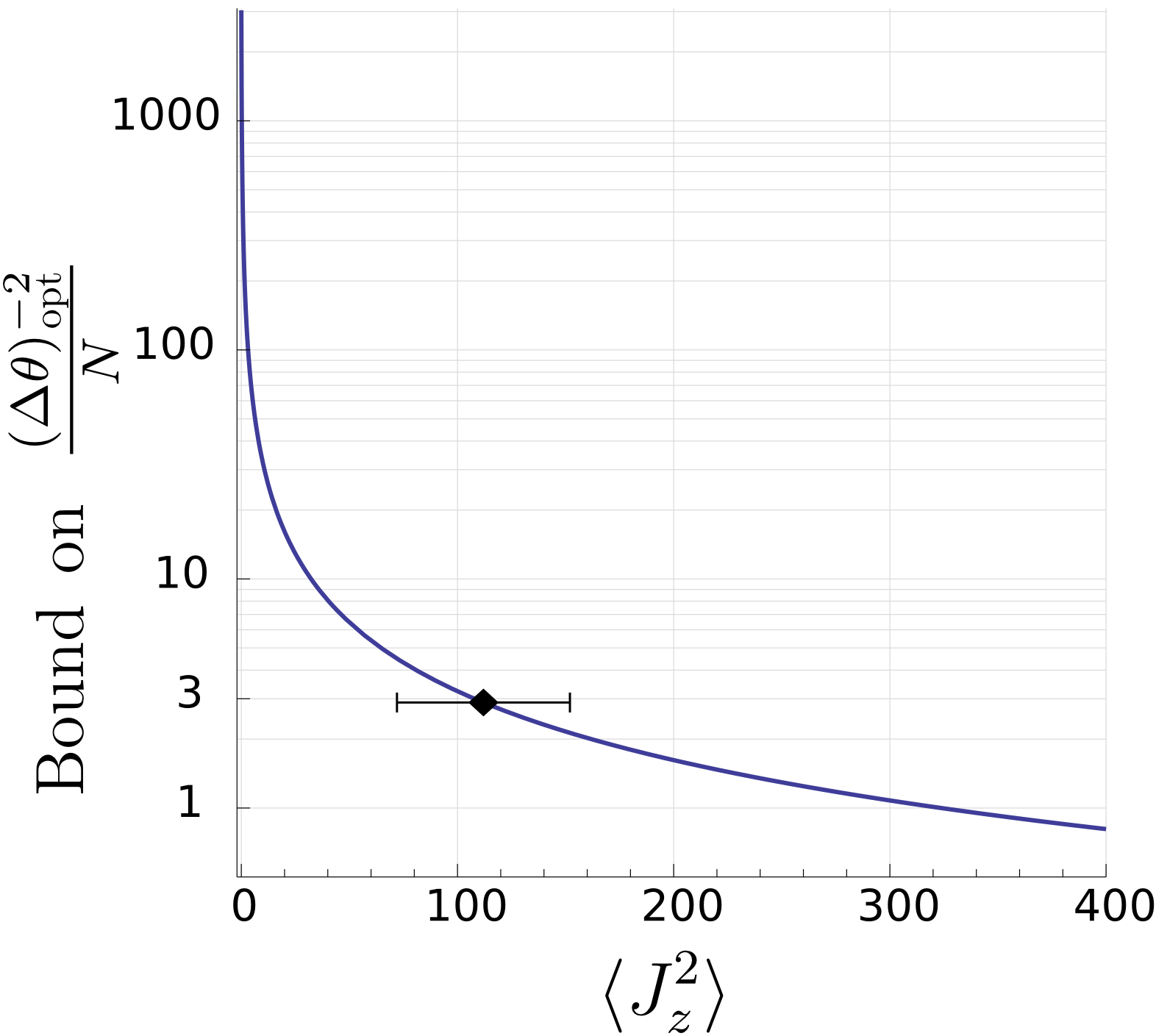} 

\vskip0.2cm
\hskip1.1cm (a) \hskip6.4cm (b)
\caption{%(Color online) 
(a)  Lower bound on  $(\Delta \theta)^{-2}$ as a function of  $\meanO{J_x^2}$ and $\meanO{J_z^2}$ for an ensemble of $N=7900$ particles.  
 For all points in a region with index $k,$ 
 the precision is bounded according to $(\Delta \theta)^{-2}>(k-1)N,$ and the state possesses at least
 $k$-particle entanglement [see \EQ{eq:FQmNF}].
 Any state corresponding to points below the curve labelled "shot-noise" is entangled.
The lower bound on $(\Delta \theta)^{-2}$ is based on \EQ{eq:estimatedeltaphi}. In addition,
  \EQ{eq:approx} is used to bound  $\meanO{J_x^4}$ and a Gaussian distribution is assumed for the measurement results of $J_z.$ 
Note that the horizontal axis is normalized by $J_{\rm max}^2$ which equals $\exs{J_x^2}$  for Dicke states given in \EQ{eq:paramvalues3}. 
The cross and the uncertainty ellipse correspond to the experimental results given in \EQ{eq:paramvalues}.
(b) A cross section corresponding to the vertical dashed line in figure (a). The uncertainty for 
the error bar is taken from  \EQ{eq:paramvalues}.}
\label{fig:2D}
\end{center}
\end{figure}

Substituting \EQ{eq:approx} and \EQ{eq:approxz} into \EQ{eq:estimatedeltaphi}, we obtain a formula that gives an upper bound on $(\Delta \theta)^{2}$ merely
as a function of $\meanO{J_x^2},$ $\meanO{J_z^2}$ and $\beta.$ 
It is reasonable to choose $\beta=3$ assuming a Gaussian statistics for the measurement results of $J_z.$
\FIGL{fig:2D}(a) shows the two-dimensional plot which is obtained
based on these considerations.
The regions with various levels of multipartite entanglement can clearly be identified.
The ideal Dicke state \eqref{eq:Dicke} corresponds to the bottom--right corner. 
In \FIG{fig:2D}(b), the cross section of the two-dimensional plot is shown.
Note that calculations based only on the second moments give a metrological usefulness different from \EQ{eq:result},
which used information also on the fourth order moments. Finally, also note that a figure similar to \FIG{fig:2D}(a)
appears in \REF{PhysRevLett.112.155304}, where multipartite entanglement has been detected independently from metrology.

%Note that calculations based only on the second moments
%underestimate somewhat the metrological usefulness of the state.
%\EQ{eq:result} gave a better bound, since it used information also on the fourth %order moments.

\section{Conclusions}
We have discussed how to access the metrological usefulness of noisy Dicke states for estimating the angle of rotation.
Our formula is able to verify the metrological usefulness without carrying out the metrological task.
We have demonstrated the use of our formula for recent experimental results.
The metrological usefulness can be inferred from measurements of  second and the fourth moments of the 
$x$-component and the $z$-component of the collective angular momentum only. In practice,
the fourth-order moments can be well approximated by the second-order moments.
After completing our calculations, we have recently become aware of a related work by Haine {\it et al.} \cite{2014arXiv1411.5111H},
which is based on the preliminary work in  \REF{2014arXiv1408.0067S},
and obtains sensitivity bounds for metrology with twin-Fock states.

\section*{Acknowledgments}

We thank S.~Altenburg, R.~Demkowicz-Dobrza\'nski, F.~Fr\"owis, O.~G\"uhne, 
P.~Hyllus, M.~Kleinmann, M.~W.~Mitchell, M.~Modugno, L.~Pezze, L.~Santos, A.~Smerzi, I.~Urizar-Lanz, and G.~Vitagliano  for stimulating discussions. We are thankful for the support of the EU
(ERC Starting Grant  258647/GEDENTQOPT, CHIST-ERA QUASAR,
the MINECO (Project No. FIS2012-36673-C03-03), the Basque
Government (Project No. IT4720-10), UPV/EHU Program No. UFI
11/55, the OTKA (Contract No. K83858).
We acknowledge support from the Centre QUEST, the DFG (Research Training Group
1729), and the EMRP, which is jointly funded by the EMRP participating countries
within EURAMET and the European Union.

\appendix

%\section*{Appendix}

\section{Details of the derivation of \EQ{eq:errorprop_den} and \EQ{eq:errorprop_num} using the symmetry 
\eqref{eq:minustheta}} 

In this appendix, we discuss how we use the symmetry \eqref{eq:minustheta} to simplify our calculations.

First, let us see the numerator of \EQ{eq:errorprop_Dicke}. Based on \EQ{eq:dyn}, the dynamics of the second and the fourth moments are obtained, respectively, as
\bea
\meanO{J_z^2 (\theta)} &=&
\meanO{J_z^2} \cos^2\theta + \meanO{J_x^2} \sin^2\theta-\exs{\{J_z,J_x\}} \sin\theta\cos\theta, \nonumber\\
\meanO{J_z^4 (\theta)} &=& \meanO{J_z^4} \cos^4\theta + 
\meanO{J_x^4}\sin^4\theta \nonumber\\ &+& 
\bigg(\meanO{\{J_z,J_x\}^2}+\meanO{\{J_z^2,J_x^2\}}\bigg) 
\cos^2\theta\sin^2\theta\nonumber\\ 
&-&\exs{A} \cos\theta\sin^3\theta - \exs{B} \cos^3\theta\sin\theta.
\label{eq:errorprop_den1}
\eea
where $A=\{J_z^2,J_xJ_z+J_zJ_x\}, $ and $B=\{J_x^2,J_xJ_z+J_zJ_x\}.$ 
%and $\{X,Y\}%=XY+XY$ is the anticommutator of $X$ and $Y.$
After calculating the terms in the numerator of \EQ{eq:errorprop_Dicke},
for the derivative in the denominator  of \EQ{eq:errorprop_Dicke} we obtain
\bea
\partial_{\theta}\meanO{J_z^2}&=&2(\meanO{J_x^2}-\meanO{J_z^2})\cos\theta\sin\theta
+\exs{\{J_z,J_x\}} (\cos^2\theta-\sin^2\theta).
\label{eq:errorprop_num1}
\eea
For calculating \EQ{eq:errorprop_num1}, we used the dynamics of of $\langle{J_z^2(\theta)}\rangle$ given in \EQ{eq:errorprop_den1}. 

Let us use now the assumption \eqref{eq:minustheta} to simplify the equations \eqref{eq:errorprop_den1}
and  \eqref{eq:errorprop_num1}.  From \eqref{eq:minustheta} it follows that the coefficients of all terms 
that are odd functions of $\theta$ must be zero. First, let us consider the equation giving the dynamics
of $\langle{J_z^2(\theta)}\rangle$ in \EQ{eq:errorprop_den1}. 
We realize that
\be
\exs{\{J_z,J_x\}} =0\label{eq:jxz0}
\ee
must hold.
Then, we set the coefficients of all other terms that are odd functions of $\theta$ to zero. Hence, from \EQ{eq:errorprop_den1}
we arrive at \eqref{eq:errorprop_den}. In a similar way, using \EQ{eq:jxz0},  
from \EQ{eq:errorprop_num1} we arrive at \EQ{eq:errorprop_num}.

Finally, as we discussed before, the experimentally prepared state has the  symmetry \eqref{eq:minustheta},
which is also an assumption used to derive \EQ{eq:mainformula}. Let us examine the case of an initial state  
$\varrho $ that does not have this property. Let us define now the state
\be
\varrho_z=\sigma_z^{\otimes N} \varrho \sigma_z^{\otimes N},\label{eq:transfz}
\ee
Direct substitution of \EQ{eq:transfz} into \EQ{eq:errorprop_den1} shows that
\bea
\langle{J_z^2(\theta)}\rangle_{\varrho}&=&\langle{J_z^2(-\theta)}\rangle_{\varrho_z},\nonumber\\
\langle{J_z^4(\theta)}\rangle_{\varrho}&=&\langle{J_z^4(-\theta)}\rangle_{\varrho_z}.
\label{eq:minustheta2}
\eea
Hence, the state
\be
\varrho_{\rm s}=\tfrac{1}{2}(\varrho+\sigma_z^{\otimes N} \varrho \sigma_z^{\otimes N} ),\label{eq:transf}
\ee
obeys the symmetry \eqref{eq:minustheta}.
Note that  the transformation \eqref{eq:transf} does not change the quantities  in \EQ{eq:mainformula2} and in \EQ{eq:mainformula}. Thus, in a sense with our scheme we get information on the metrological usefulness
of the state $\varrho_{\rm s},$ that we would get after the trivial mixing operation  \eqref{eq:transf}.

\section{ Details of the calculations for \EQ{eq:mainformula2}}

\commentcolor{
First, we compute the six fourth-order moments appearing as coefficients of
the $\cos^2\theta\sin^2\theta$ in  \EQ{eq:errorprop_den}, after expanding the anticommutators
\bea
\meanO{J_zJ_xJ_zJ_x}&=&i\meanO{J_zJ_xJ_y}+\meanO{J_zJ_x^2J_z},\nonumber\\
\meanO{J_zJ_x^2J_z}&=&\meanO{J_zJ_x^2J_z},\nonumber\\
\meanO{J_xJ_zJ_xJ_z}&=&-i\meanO{J_yJ_xJ_z}+\meanO{J_zJ_x^2J_z},\nonumber\\
\meanO{J_xJ_z^2J_x}&=&-i\meanO{J_yJ_zJ_x}+i\meanO{J_zJ_xJ_y}+
\meanO{J_zJ_x^2J_z},\nonumber\\
\meanO{J_z^2J_x^2}&=&i\meanO{J_zJ_yJ_x}+i\meanO{J_zJ_xJ_y}
+\meanO{J_zJ_x^2J_z},\nonumber\\
\meanO{J_x^2J_z^2}&=&-i\meanO{J_xJ_yJ_z}-i\meanO{J_yJ_xJ_z}+
\meanO{J_zJ_x^2J_z},\label{eq:A1}
\eea
where simple commutation relations are used. If now we sum all the right-hand sides of the equations \eqref{eq:A1}, we obtain the sum of
$6\meanO{J_zJ_x^2J_z}$ and a term with third-order moments  with $J_x,$ $J_y,$ and $J_z.$
Hence, we obtain
\be
\meanO{\{J_z,J_x\}^2}+\meanO{\{J_z^2,J_x^2\}}=6\meanO{J_zJ_x^2J_z}+\mathcal{T},
\ee
where 
\bea
\mathcal{T}&:=&i(\meanO{J_zJ_xJ_y}-\meanO{J_yJ_xJ_z}-\meanO{J_yJ_zJ_x}+\meanO{J_zJ_xJ_y}\nonumber\\
&+&\meanO{J_zJ_yJ_x}+\meanO{J_zJ_xJ_y}-\meanO{J_xJ_yJ_z}-\meanO{J_yJ_xJ_z})=
\nonumber\\
&=&i(\meanO{[J_z,J_xJ_y]}+\meanO{[J_z,J_yJ_x]}+\meanO{[J_zJ_x,J_y]}
+\meanO{[J_z,J_xJ_y]}+\meanO{[J_x,J_y]J_z})
\nonumber\\
&=&4\meanO{J_x^2}-3\meanO{J_y^2}-2\meanO{J_z^2}. \label{eq:333}
\eea}

In this appendix, we give further details of our calculations for obtaining \EQ{eq:mainformula2}.
After straightforward but long algebra based on commutation relations,
the coefficient of
the term $\cos^2\theta\sin^2\theta$ in  \EQ{eq:errorprop_den} can be obtained as
\be\label{eq:coeffsincos}
\meanO{\{J_z,J_x\}^2}+\meanO{\{J_z^2,J_x^2\}}=4\meanO{J_x^2}-3\meanO{J_y^2}-2\meanO{J_z^2}+6\exs{J_zJ_x^2J_z}.
\ee
Then, based on \EQ{eq:errorprop_Dicke}, \EQ{eq:errorprop_den}, \EQ{eq:errorprop_num}, and \EQ{eq:coeffsincos}, we can write the variance of the estimated parameter $\theta$ in the following way
\be\label{eq:deltathetafraction}
(\Delta \theta)^{2}=\frac
{(\Delta J_z^2)^2\cos^4\theta+(\Delta J_x^2)^2\sin^4\theta
+\mathcal{C}\cos^2\theta\sin^2\theta}{4(\meanO{J_x^2}-\meanO{J_z^2})^2
\cos^2\theta\sin^2\theta},
\ee 
where
\be\label{eq:C}
\mathcal{C}:=4\meanO{J_x^2}-3\meanO{J_y^2}-2\meanO{J_z^2}+6\exs{J_zJ_x^2J_z}-2\meanO{J_x^2}\meanO{J_z^2}.
\ee
%In \EQ{eq:C} we can recognize the terms coming from  \EQ{eq:deltathetafraction}.
Simplifying and rearranging terms in \EQ{eq:deltathetafraction}, we arrive at \EQ{eq:mainformula2}. Note that without the assumption
\eqref{eq:minustheta},   we could not have obtained a 
formula with so simple dependence on $\theta.$

\vskip 15pt

\bibliography{dickemetro}

\end{document}